# Crash Severity Pattern of Motorcycle Crashes in Developing Country Context

**Sina Asgharpour, Corresponding Author**
Ph.D. Candidate
Department of Civil, Materials, and Environmental Engineering
University of Illinois at Chicago
842 W Taylor St, ERF, Chicago, IL 60607
Email: sasgha3@uic.edu

**Mohammadjavad Javadinasr**
Ph.D. Candidate
Department of Civil, Materials, and Environmental Engineering
University of Illinois at Chicago
842 W Taylor St, ERF, Chicago, IL 60607
Email: mjavad2@uic.edu

**Zeinab Bayati**
M.Sc. Student
School of Engineering
College of Computing, Engineering, and Construction
University of North Florida
1 UNF Drive, Skinner - Jones Hall, Building 4 STE. 4201, Jacksonville, FL 32224
Email: Zeinab.bayati@unf.edu

**Abolfazl (Kouros) Mohammadian**
Professor
Department of Civil, Materials, and Environmental Engineering
University of Illinois at Chicago
842 W Taylor St, ERF, Chicago, IL 60607
Email: kouros@uic.edu

Word count: 6,047 words + 5 tables (250 words per table) = 7,297 words

*Submitted August 1, 2021*

**ABSTRACT**

Despite paying special attention to the motorcycle-involved crashes in the safety research, little is known about their pattern and impacts in developing countries. The widespread adoption of motorcycles in such regions in tandem with the vulnerability of motorcyclists exacerbates the likelihood of severe crashes. The main objective of this paper is to investigate the underlying factors contributing to the severity of motorcycle-involved crashes through employing crash data from March 2018 to March 2019 from Iran. Considering the ordinal nature of three injury classes of property-damage-only (PDO), injury, and fatal crashes in our data, an ordered logistic regression model is employed to address the problem. The data statistics suggest that motorcycle is responsible for 38% of injury and 15% of all fatal crashes in the dataset. The results indicate that significant factors contributing to more severe crashes include collision, road, temporal, and motorcycle rider characteristics. Among all attributes, our model is most sensitive to the motorcycle-pedestrian accident, which increases the probability of belonging a crash into injury and fatal crashes by 0.289 and 0.019, respectively. Moreover, we discovered a significant degree of correlation between young riders and riders without a license. Finally, upon the insights obtained from the results, we propose safety countermeasures, including 1) strict traffic rule enforcement upon riders and pedestrians, 2) educational programs, and 3) road-specific adjustment policies.

# 1 INTRODUCTION

Due to lack of protective structure, motorcycles are one of the most vulnerable vehicles exposed to high injury and fatality risk. More than half of the road crash fatalities are among vulnerable road users (i.e., motorcyclists, pedestrians, and bicyclists) (*1*). Moreover, per vehicle miles traveled in 2019 in the U.S., motorcyclists' risk to be involved in a fatal or injury crash is 29 and 4 times more than a car, respectively(*2*).

The severity of motorcycle-involved crashes in developing countries is also becoming a major concern, primarily for the growing use of motorcycles as an affordable transportation mode (*3*). For instance, the fraction of motorcycles to all motorized vehicles is 55% for Sri Lanka (per capita income of $3,780) (*1*), while it is nearly 3% for the U.S. ($56,180 per capita income) (*1*). These high ratios of adoption underscore the importance of analyzing the pattern of motorcycle-involved crashes in developing countries; yet, the research in this area remains sparse (*3*).

Similar to other developing countries, motorcycles are a popular transportation mode in Iran with a motorcycles-to-all vehicles ratio of 38% (with per capita income of $6,530) (*1*). Also, there are 3.5 million motorcycles in Tehran (i.e., the capital of Iran), among which 600,000 commute daily (*4*). Despite their crucial importance, studies that have addressed the severity of motorcycle-involved crashes in Iran are rare or limited to a specific area (*5*).

This study aims to investigate factors affecting the severity of motorcycle-involved crashes in Iran. To this end, we utilize the ordered logit model and crash records of Iran from March 2018 to March 2019 to estimate the parameters of the model. In addition, employing statistical analyses, we determine the significant patterns between the variables to better understand the results of the model. The findings of this study can provide better insights into motorcycle-involved crashes to policymakers and planners. Because implementing a countermeasure without acknowledging the effect of different factors and the interplay between key factors would lead to inefficient or unstable solutions to improve the crash severity.

The remainder of this paper is organized as follows: in section 2, we review the related literature. Statistics of the crash data are presented in section 3. The details of the methodology are described in section 4. The results of the model and discussion are presented in section 5. In section 6, we elaborate on policy recommendations. Finally, the summary of the study is presented in section 7.

## 2 BACKGROUND

The severity of motorcycle-involved crashes has an extensive research background and considerable efforts have been made to explore the key influential factors in recent years. Among collision characteristics, single and multi-vehicle motorcycle crashes are widely examined and found to be significant (*6–10*). In this regard, motorcycle-pedestrian (*11–14*) and stationary object collisions (*8, 15, 16*) are usually correlated with severe crashes. Moreover, road type (*7, 16, 17*) and road geometry (*6, 19*) (e.g., horizontal/vertical curves) are also two paramount factors. Special ambient conditions, such as light level (*9*), motorcyclist conspicuity (*20*), and surface condition (*21*), are recognized as factors influencing crash severity. Regarding the attributes of a motorcycle rider, young riders are one of the most vulnerable groups, mainly for lack of experience (*6, 17, 22*). Furthermore, human factors, such as helmet use (*8*), alcohol or drug impairment (*8*), and speeding behavior (*23*), are of special interest. As a summary, **Table 1** presents the significant explanatory variables in the reviewed studies of motorcycle accidents from 2003 to 2021.

Various modeling tools are used to assess the impact of different variables on motorcycle crash severity. Lin et al. (*15*) utilized the ordered logit model to ascertain risk factors associated with injury among young riders. Shaheed and Gkritza (*8*) used a latent class multinomial logit model to identify dominant severity characteristics of single-vehicle motorcycle crashes. Savolainen and Mannering (*10*) employed nested and multinomial logit models to assess the contributing factors to the severity of single and multi-vehicle crashes. In addition, Wahab and Jiang (*24*) utilized machine learning algorithms to investigate risk factors impacting injury severity of motorcycle crashes. Furthermore, Halbersberg and Lerner (*22*) determined factors associated with fatal crashes of young motorcycle riders using the Bayesian network.

Motorcycle crashes in developing countries have some distinct characteristics that challenge utilizing policies implemented in developed countries to improve motorcycle-related safety. Lin and Kraus (*3*) identified some of these characteristics, including 1) rapid growth of motorcycle use, 2) roadway environment features (e.g., traffic congestion), 3) poor traffic rules enforcement or inadequate education. Pervez et al. (*25*) analyzed crash data from Karachi city using a random parameter logit model. According to the result, they suggested strict enforcement measures to control motorcyclists' risky behavior and exclusive motorcycles lanes. Additionally, Manan et al. (*7*) determined risk factors associated with motorcycle single and multi-vehicle fatal crashes in Malaysia and recommend exclusive lanes for motorcyclists. **Table 2** summarizes the key components of a subset of studies focusing on developing countries, including their data features, analysis tools, and major findings. Despite the studies mentioned above, few works investigate motorcycle crash severity in developing countries compared to developed ones (*3*).

**Table 1 Significant Explanatory Variables in the Reviewed Studies of Motorcycle Accidents from 2003 to 2021**

| Studies (Sorted by Year in Descending Order) | Collision Type | Collision Mode | Road Type | Road Geometry | Land-use | Human Factor | Time of Day | Season | Weather | Surface Condition | Rider's Fault | Rider's Gender | Rider's Age | Rider's Education | Rider's License |
|---|---|---|---|---|---|---|---|---|---|---|---|---|---|---|---|
| Pervez et al. (*25*) | ✓ | | | | | | ✓ | ✓ | | | | ✓ | ✓ | | |
| Farid and Ksaibati (*6*) | ✓ | | ✓ | ✓ | | | ✓ | | | ✓ | | ✓ | ✓ | | |
| Vajari et al. (*26*) | ✓ | | | | ✓ | | ✓ | | ✓ | ✓ | | ✓ | ✓ | | |
| Shajith et al. (*27*) | ✓ | ✓ | | | | ✓ | ✓ | | | ✓ | | ✓ | ✓ | | |
| Waseem et al. (*16*) | ✓ | ✓ | ✓ | | | | ✓ | ✓ | ✓ | | | ✓ | ✓ | ✓ | |
| Quddus et al. (*11*) | ✓ | | ✓ | | | | ✓ | ✓ | | ✓ | | ✓ | ✓ | | |
| Halbersberg and Lerner (*22*) | ✓ | ✓ | ✓ | | ✓ | | ✓ | | | | | ✓ | ✓ | | |
| Wahab and Jiang (*24*) | ✓ | ✓ | | ✓ | ✓ | | ✓ | | | ✓ | | | | | |
| Manan et al. (*7*) | | | ✓ | ✓ | ✓ | | ✓ | | | ✓ | ✓ | | | | |
| Das et al. (*17*) | ✓ | ✓ | ✓ | ✓ | | | ✓ | | | | ✓ | | | | |
| Oliveira et al. (*28*) | | | | | | | | | | | | ✓ | ✓ | ✓ | ✓ |
| Chichom-Mefire et al. (*29*) | ✓ | | | | | | ✓ | | | | | ✓ | ✓ | | |
| Shaheed and Gkritza (*8*) | ✓ | | | | | ✓ | ✓ | ✓ | ✓ | | ✓ | ✓ | ✓ | | |
| Jones et al. (*19*) | | ✓ | | ✓ | ✓ | ✓ | ✓ | | | ✓ | | ✓ | ✓ | | |
| Rifaat et al. (*30*) | | ✓ | ✓ | | | ✓ | ✓ | ✓ | | | | ✓ | | | |
| Haque et al. (*21*) | | | | | | | ✓ | | | ✓ | | | | | |
| Geedipally et al. (*9*) | ✓ | | ✓ | ✓ | ✓ | ✓ | ✓ | | | | ✓ | ✓ | ✓ | | |
| Eustace et al. (*31*) | | ✓ | | ✓ | ✓ | ✓ | ✓ | ✓ | ✓ | | | ✓ | ✓ | | |
| Oginni et al. (*32*) | ✓ | ✓ | | | | | | | | | | ✓ | ✓ | | |
| Savolainen and Mannering (*10*) | | ✓ | | ✓ | | | ✓ | | | ✓ | ✓ | ✓ | ✓ | | |
| Chang and Yeh (*33*) | ✓ | | ✓ | ✓ | | ✓ | ✓ | | | | | ✓ | ✓ | | |
| Lin et al. (*15*) | ✓ | | ✓ | ✓ | ✓ | ✓ | ✓ | | ✓ | ✓ | | ✓ | ✓ | | ✓ |
| Iamtrakul et al. (*34*) | | | | | | ✓ | ✓ | ✓ | ✓ | | | ✓ | ✓ | | |

**Table 2 Reviewed Motorcycle Crash Severity Studies in Developing Countries**

| Studies (Sorted by Year in Descending Order) | Country | Data Year (Observations) | Method | Main Contributing Factors to Severe Crashes |
|---|---|---|---|---|
| Pervez et al. (*25*) | Pakistan | 2014-2015 (28,894) | Random Parameter Logit Model | summer season, weekends, night-time, heavy and single-vehicle collisions, young and elderly riders, and female pillions (clothes stuck in the wheel) |
| Shajith et al. (*27*) | Sri Lanka | 2012-2014 (N/A) | Binary Logit Model | wet surface condition, night-time, elderly riders |
| Quddus et al. (*11*) | Singapore | 1992-2000 (27,570) | Ordered Probit Model | stationary objects and pedestrians collisions, engine capacity of the motorcycle, young riders, and male riders' risky driving behavior |
| Wahab and Jiang (*24*) | Ghana | 2011-2015 (8,516) | Machine Learning | the day of the week, time of the day, collision type, road separation, road surface type, and road shoulder condition |
| Manan et al. (*7*) | Malaysia | 2010-2012 (9,176) | Multinomial and Mixed Logit Models | for single-vehicle crashes: roadway curves, no road marking, slippery surface<br>for multi-vehicle crashes: expressway, primary and secondary roads, speed limit more than 70 km/h |
| Haque et al. (*21*) | Singapore | 2004-2008 (21,922) | Log-linear Models | night-time, wet road surfaces, and road side conflicts |
| Lin et al. (*15*) | Taiwan | 1994-1996 (4,729) | Ordered Logit Model | rural road, heavier object collisions, darkness, and speeding |

## 3 DATA

In this section, specifications of the data used in this study are presented. Our data contain nationwide crash information of Iran recorded by traffic authorities from March 2018 to March 2019. Information of crash injury severity is recorded in three categories: Property-damage only (PDO), injury (non-fatal injury), and fatal. Injury or fatal records refer to crashes in which at least one occupant involved in the crash is injured or dead, respectively.

In the following subsections, firstly, we demonstrate the significance of motorcycle crash severity compared to other crashes using the dataset of all crashes in the country. Then, we present descriptive statistics of motorcycle-involved crashes, which is a subset of all crashes.

### 3.1 Significance of Motorcycle-related Severity

As illustrated in **Figure 1**, 20% of all accidents are motorcycle-involved. Yet, motorcycle-involved crashes account for 5% of PDO (12,737 out of 280,948), 38% of injury (93,013 out of 244,659), and 15% of fatal (1,197 out of 7,802) crashes. Moreover, 87% of motorcycle-involved crashes involve injury, whereas this fraction for other vehicles is 28.4% signifying the remarkable share of motorcycle-caused injury.

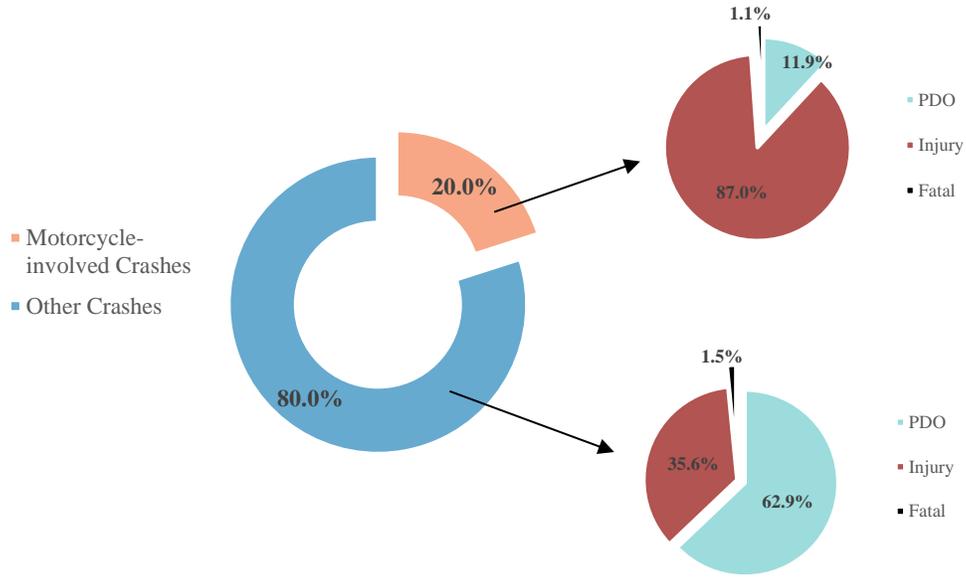

**Figure 1: Distribution of Motorcycle-involved vs. Other Crashes and Their Corresponding Share of PDO, Injury, and Fatal Severity Classes**

To further elaborate on the degree of association between motorcycle-involved and injury crashes, we performed Pearson's Chi-square test (*35*) according to Eq. (1) and (2), where $\chi^2$ is the chi-square statistics, $res_{r,c}$ is the residual value for cell $r, c$ (i.e., row $r$ and column $c$), $E_{r,s}$ and $O_{r,c}$ represent the estimated and observed frequencies for cell $r, c$.

$$\chi^2 = \sum_{r,c} \frac{\left(E_{r,c} - O_{r,c}\right)^2}{E_{r,c}} \tag{1}$$

$$res_{r,c} = \frac{E_{r,c} - O_{r,c}}{\sqrt{E_{r,c}}} \ , \qquad \forall r, c \tag{2}$$

Results of the chi-square test between vehicle type and severity category indicate the significant correlation between these variables. **Figure 2** depicts the associated residual values as well as frequency percentage of each cell, suggesting that the most degree of the association is captured by motorcycle-involved crashes. The residuals of the motorcycle category show a significant positive correlation of this category with injury crashes and a negative correlation with PDO.

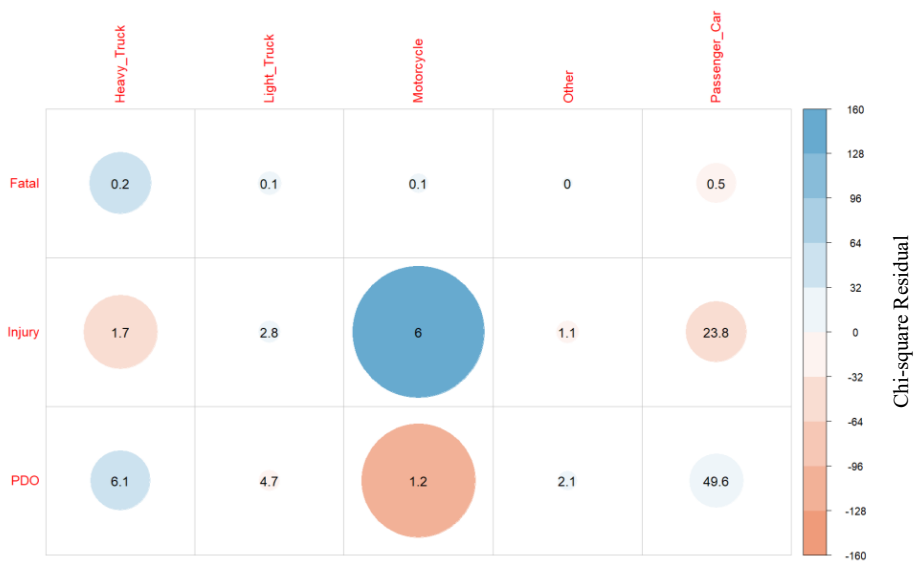

**Figure 2: Chi-Square Test Residual between Crash Severity and Vehicle Types** (Size of bubbles is proportional to the residuals and numbers in cells signifies frequency % of them)

### 3.1 General Statistics of the Motorcycle Dataset

**Table 3** shows the frequency of variables of the motorcycle-involved crashes cross-tabulated with the injury classes. Explanatory variables of the model are a subset of the variables provided in **Table 3.**

**Table 3 Descriptive Statistics of Motorcycle-involved Crashes**

| Variable | | PDO | | Injury | | Fatal | | Total | |
|---|---|---|---|---|---|---|---|---|---|
| | Total | 6,486 | 16.7% | 31,926 | 82.2% | 421 | 1.1% | 38,833 | 100.0% |
| Collision Type | Motor-Vehicle | 5,441 | 18.3% | 24,037 | 80.9% | 240 | 0.8% | 29,718 | 76.5% |
| | Two-Motor | 506 | 16.0% | 2,599 | 82.3% | 53 | 1.7% | 3,158 | 8.1% |
| | Motor-Pedestrian | 6 | 0.3% | 2,258 | 98.5% | 29 | 1.3% | 2,293 | 5.9% |
| | Overturn | 92 | 5.5% | 1,550 | 92.1% | 41 | 2.4% | 1,683 | 4.3% |
| | Multi-Vehicle (>2) | 243 | 37.6% | 389 | 60.2% | 14 | 2.2% | 646 | 1.7% |
| | Multiple | 136 | 28.0% | 344 | 70.8% | 6 | 1.2% | 486 | 1.3% |
| | Other | 62 | 7.3% | 749 | 88.2% | 38 | 4.5% | 849 | 2.2% |
| Collision Mode | Front-Left | 1,057 | 13.8% | 6,569 | 85.6% | 44 | 0.6% | 7,670 | 19.8% |
| | Front-Right | 990 | 14.1% | 5,972 | 85.1% | 54 | 0.8% | 7,016 | 18.1% |
| | Front-End (Head-on) | 1,066 | 16.1% | 5,451 | 82.2% | 117 | 1.8% | 6,634 | 17.1% |
| | Front-Rear | 1,473 | 26.2% | 4,083 | 72.5% | 74 | 1.3% | 5,630 | 14.5% |
| | Right-Left | 962 | 20.7% | 3,672 | 79.0% | 16 | 0.3% | 4,650 | 12.0% |
| | Other | 938 | 13.0% | 6,179 | 85.4% | 116 | 1.6% | 7,233 | 18.6% |
| Road Type | Main Street | 4,299 | 16.6% | 21,525 | 82.9% | 136 | 0.5% | 25,960 | 66.9% |
| | Secondary Street | 485 | 14.0% | 2,970 | 85.4% | 21 | 0.6% | 3,476 | 9.0% |
| | Main Road | 506 | 16.1% | 2,554 | 81.1% | 89 | 2.8% | 3,149 | 8.1% |
| | Highway | 602 | 26.8% | 1,598 | 71.1% | 47 | 2.1% | 2,247 | 5.8% |
| | Secondary Road | 313 | 14.9% | 1,714 | 81.6% | 74 | 3.5% | 2,101 | 5.4% |
| | Rural Road | 196 | 14.4% | 1,120 | 82.3% | 45 | 3.3% | 1,361 | 3.5% |
| | Other | 85 | 15.8% | 445 | 82.6% | 9 | 1.7% | 539 | 1.4% |
| Road Geometry | Flat | 6,195 | 16.7% | 30,542 | 82.3% | 368 | 1.0% | 37,105 | 95.6% |
| | Horiz./Vert. Curve | 291 | 16.8% | 1,384 | 80.1% | 53 | 3.1% | 1,728 | 4.4% |
| Land-use | Residential | 5,181 | 17.5% | 24,240 | 81.9% | 190 | 0.6% | 29,611 | 76.3% |
| | Not Residential | 1,305 | 14.2% | 7,686 | 83.3% | 231 | 2.5% | 9,222 | 23.7% |
| Human Factor | Reckless/Hurry | 4,595 | 19.6% | 18,592 | 79.5% | 213 | 0.9% | 23,400 | 60.3% |
| | None | 1,563 | 13.9% | 9,508 | 84.8% | 137 | 1.2% | 11,208 | 28.9% |
| | Rule Ignorance | 193 | 6.1% | 2,919 | 92.4% | 46 | 1.5% | 3,158 | 8.1% |
| | Other | 135 | 12.7% | 907 | 85.0% | 25 | 2.3% | 1,067 | 2.7% |
| Season | Spring/Summer | 3,276 | 14.3% | 19,404 | 84.6% | 263 | 1.1% | 22,943 | 59.1% |
| | Fall/Winter | 3,210 | 20.2% | 12,522 | 78.8% | 158 | 1.0% | 15,890 | 40.9% |
| Time of Day | Day | 4,892 | 18.5% | 21,371 | 80.7% | 209 | 0.8% | 26,472 | 68.2% |
| | Night | 1,409 | 12.6% | 9,573 | 85.7% | 191 | 1.7% | 11,173 | 28.8% |
| | Down/Dusk | 185 | 15.6% | 982 | 82.7% | 21 | 1.8% | 1,188 | 3.1% |
| Weather | Clear | 6,319 | 16.7% | 31,186 | 82.3% | 402 | 1.1% | 37,907 | 97.6% |
| | Unclear | 167 | 18.0% | 740 | 79.9% | 19 | 2.1% | 926 | 2.4% |
| Surface Condition | Dry | 6,380 | 16.7% | 31,474 | 82.2% | 414 | 1.1% | 38,268 | 98.5% |
| | Wet/Icy/Muddy | 106 | 18.8% | 452 | 80.0% | 7 | 1.2% | 565 | 1.5% |
| Rider's Fault | not at-Fault | 2,815 | 12.2% | 20,087 | 87.0% | 177 | 0.8% | 23,079 | 59.4% |
| | at-Fault | 3,671 | 23.3% | 11,839 | 75.1% | 244 | 1.5% | 15,754 | 40.6% |
| Rider's Gender | Male | 6,474 | 16.7% | 31,868 | 82.2% | 420 | 1.1% | 38,762 | 99.8% |
| | Female | 12 | 16.9% | 58 | 81.7% | 1 | 1.4% | 71 | 0.2% |

**Table 3 (Continued)**

| Variable | | PDO | | Injury | | Fatal | | Total | |
|---|---|---|---|---|---|---|---|---|---|
| Rider's Age | Under25 | 1,740 | 12.3% | 12,182 | 86.4% | 176 | 1.2% | 14,098 | 36.3% |
| | 26to55 | 4,480 | 19.9% | 17,789 | 79.2% | 197 | 0.9% | 22,466 | 57.9% |
| | Over56 | 266 | 11.7% | 1,955 | 86.2% | 48 | 2.1% | 2,269 | 5.8% |
| Rider's Education | Sec. School/Less | 1,141 | 14.6% | 6,559 | 84.1% | 97 | 1.2% | 7,797 | 20.1% |
| | Diploma | 4,852 | 16.6% | 24,142 | 82.4% | 316 | 1.1% | 29,310 | 75.5% |
| | College/Higher | 493 | 28.6% | 1,225 | 71.0% | 8 | 0.5% | 1,726 | 4.4% |
| Rider's License Status | Motorcycle License | 4,318 | 30.8% | 9,664 | 69.0% | 30 | 0.2% | 14,012 | 36.1% |
| | Not Reported | 1,761 | 11.8% | 12,835 | 86.3% | 280 | 1.9% | 14,876 | 38.3% |
| | Without License | 407 | 4.1% | 9,427 | 94.8% | 111 | 1.1% | 9,945 | 25.6% |

## 4 METHOD

In this section, we elaborate on the ordered logit model as the employed methodology in this study. As described in the previous section, we use a reported crash data including 3 ordinal classes of severity. Accordingly, we utilized the ordered logit model (*30, 36, 37*) to proceed with our analysis. Rifaat et al. (*30*) utilized the employed logit model, heterogeneous choice model, and partially constrained generalized ordered logit model to analyze the severity of motorcycle-involved crashes. Per the results of their study, these models produce very similar estimates suggesting the robustness of the models.

The ordered models consider a common utility function across ordinal categories/classes which can be written as Eq. (3), where $i$ represents the observation indicator, $U_i$ is the utility function of observation $i$, $\epsilon_i$ captures the unobserved factors of observation $i$, and $\mathbf{X_i}$ and $\boldsymbol{\beta}$ are the exogenous variable vector and parameter vector, respectively. In the ordered models, there are $J - 1$ cut-off points separating $J$ categories. According to Eq. (4), the probability of being classified into category $j$ equals to the probability of $U_i$ being between two adjacent cut-off points (i.e., $C_j$ and $C_{j-1}$); where $y_i$ is the discrete category of observation $i$.

$$U_i = \mathbf{X_i} \cdot \boldsymbol{\beta} + \epsilon_i \qquad\qquad , \forall i = 1, 2, \dots, N \qquad\qquad (3)$$

$$\Pr(y_i = j) = Pr(C_{j-1} < U_i < C_j) \qquad\qquad , \forall i = 1, 2, \dots, N, \ \ \forall = 1, 2, \dots, J \qquad (4)$$

According to Eq. (3) and (4), $\Pr(y_i = j)$ takes the form shown in Eq. (5) in terms of $\epsilon_i$, the random variable signifying the unobserved factors. As a solution to Eq. (5), the ordered logit model assumes IID extreme value distribution for $\epsilon_i$ (*38*). Consequently, difference of two variables distributed as IID extreme value follows a logistic distribution. Therefore, $\Pr(y_i = j)$ can be obtained by Eq. (6) in which parameters of $\boldsymbol{\beta}$ vector and cut-off points are calculated through the model estimation process.

$$\Pr(y_i = j) = Pr\big(\epsilon_i < C_j - \boldsymbol{X_i} \cdot \boldsymbol{\beta}\big) - Pr\big(\epsilon_i < C_{j-1} - \boldsymbol{X_i} \cdot \boldsymbol{\beta}\big) \qquad , \forall i = 1, 2, \dots, N, \;\; \forall j = 1, 2, \dots, J \quad (5)$$

$$\Pr(y_i = j) = \frac{\exp\left(C_j - \boldsymbol{X_i} \cdot \boldsymbol{\beta}\right)}{1 + \exp\left(C_j - \boldsymbol{X_i} \cdot \boldsymbol{\beta}\right)} - \frac{\exp\left(C_{j-1} - \boldsymbol{X_i} \cdot \boldsymbol{\beta}\right)}{1 + \exp\left(C_{j-1} - \boldsymbol{X_i} \cdot \boldsymbol{\beta}\right)} \qquad , \forall i = 1, 2, \dots, N, \;\; \forall j = 1, 2, \dots, J \quad (6)$$

## 5  RESULTS

As mentioned in section 3, we considered three injury classes, including PDO, injury, and fatal. As a part of the model specification procedure, we arranged these classes in an increasing severity order in the ordered logit model. Accordingly, a positive sign of an estimated coefficient signifies the correlation of the associated variable with more severe classes. We employed Stata MP ver. 15.1.629 software to estimate coefficients of the ordered logit model through the Maximum likelihood method. Eq. (7) and Eq. (8) show the likelihood ratio chi-square statistics ($\chi^2_{LL}$) and the goodness of fit ratio ($\rho^2$), respectively. Chi-square test suggests that the model is statistically significant in general.

$$\chi^2_{LL} = 8{,}578.3 \;, \qquad \text{df} = 14 \;, \qquad \text{p} < 0.0001 \tag{7}$$

$$\rho^2 = 0.217 \tag{8}$$

**Table 4** presents the estimated coefficients of the model, associated standard errors, and t-test significance. As shown, the parameters -0.357 and 6.348 are the estimated cut-off points of the ordered logit model for PDO−injury, and injury−fatal boundaries, respectively. In addition, the statistical significance of these thresholds confirms the distinct difference between severity classes. In the following subsections, we interpret the estimation results presented in **Table 4** coupled with **Table 5**, demonstrating the marginal effect of the model parameters on the severity classes.

**Table 4 Estimation Results of the Ordered Logit Model**

| Variable/Parameter | | Estimated Coefficient | Standard Error | t statistics |
|---|---|---|---|---|
| Thresholds | Cut-off Point 1 | -0.357 *** | 0.032 | 10.31 |
| | Cut-off Point 2 | 6.348 *** | 0.066 | -95.93 |
| Collision Type | Motor-Pedestrian | 2.553 *** | 0.087 | 29.40 |
| | Overturn | 2.007 *** | 0.094 | 21.44 |
| | Two-Motor | 0.391 *** | 0.052 | 7.49 |
| Collision Mode | Head-on | 0.087 ** | 0.038 | 2.31 |
| Road Type | Highway | -0.347 *** | 0.053 | -6.51 |
| | Secondary Street | 0.091 * | 0.050 | 1.82 |
| Season | Spring/Summer | 0.391 *** | 0.028 | 14.12 |
| Time of Day | Night | 0.381 *** | 0.033 | 11.66 |
| | Dawn/Dusk | 0.179 ** | 0.083 | 2.16 |
| Rider's Fault | not at-Fault | 1.078 *** | 0.029 | 37.36 |
| Rider's Age | Under 25 | 0.490 *** | 0.031 | 16.06 |
| | Over 56 | 0.743 *** | 0.067 | 11.13 |
| Rider's Education | Secondary School or Less | 0.154 *** | 0.036 | 4.30 |
| | College or Higher | -0.627 *** | 0.058 | -10.77 |

* 90% significance level,          ** 95% significance level,          *** 99% significance level

**Table 5 Marginal Effects of Estimated Coefficients**

| Variables | | Severity Class | | |
|---|---|---|---|---|
| | | PDO | Injury | Fatal |
| Collision Type | Motor-Pedestrian | -0.309 | 0.289 | 0.019 |
| | Overturn | -0.243 | 0.227 | 0.015 |
| | Two-Motor | -0.047 | 0.044 | 0.003 |
| | | | | |
| Collision Mode | Head-on | -0.011 | 0.010 | 0.001 |
| | | | | |
| Road Type | Highway | 0.042 | -0.039 | -0.003 |
| | Secondary Street | -0.011 | 0.010 | 0.001 |
| | | | | |
| Season | Spring/Summer | -0.047 | 0.044 | 0.003 |
| | | | | |
| Time of Day | Night | -0.046 | 0.043 | 0.003 |
| | Dawn/Dusk | -0.022 | 0.020 | 0.001 |
| | | | | |
| Rider's Fault | not at Fault | -0.130 | 0.122 | 0.008 |
| | | | | |
| Rider's Age | Under 25 | -0.059 | 0.055 | 0.004 |
| | Over 56 | -0.090 | 0.084 | 0.006 |
| | | | | |
| Rider's Education | Secondary School or Less | -0.019 | 0.017 | 0.001 |
| | College or Higher | 0.076 | -0.071 | -0.005 |

## 5.1 Collision Characteristics

Collision characteristics are represented by collision type and collision mode (i.e., collision angle) attributes in the estimated model. As shown in **Table 4**, three collision type variables (i.e., motor-pedestrian, overturn, and two motorcycles), and a collision mode attribute (i.e., head-on) turned out to be significant.

Per the results, motor-pedestrian crashes tend to be more severe since their associated coefficient is 2.553. According to **Table 5**, involving in a motor-pedestrian accident decreases the probability of the accident being PDO by 0.309, and increases the likelihood of injury and fatality by 0.289 and 0.019, respectively. This outcome is also in line with the literature as specified in Shajith et al. (*27*); Quddus et al. (*11*); Chen and Fan (*13*). Even in minor motorcycle-pedestrian crashes, both sides of the accident (i.e., the rider and the pedestrian) are highly vulnerable which makes them susceptible to higher risks. In addition, the lack of traffic rule regulations in Iran, especially for the pedestrians, is a serious underlying factor contributing to more severe accidents. For instance, no rigorous traffic rules are imposed on pedestrians and motorcycle riders crossing a red light. Also, in congested areas, some riders prefer to take sidewalks instead of the main road, yet there are not enough regulations tackling this risky behavior. This is also in good agreement with Shajith et al. (*27*) who acknowledged the motorcycle-pedestrian crashes on sidewalks as a noticeable factor affecting crash severity in Sri Lanka.

Overturn Crashes are the other collision type that contributes to increasing the odds of more severity and usually occur as a result of rider's loss of control in critical situations (e.g., slippery surface condition, attempt to avoid hitting another vehicle/pedestrian). Falling into this crash category raises the probability of injury and fatality by 0.227 and 0.015, respectively. This finding is also well in line with the similar findings of Pervez et al. (*25*), and Shaheed and Gkritza (*8*).

Furthermore, being involved in two-motorcycle crashes increases the likelihood of fatal and non-fatal injuries by 0.003 and 0.044, respectively. An underlying reason for this outcome could be the high vulnerability of both motorcycle riders, especially at higher speeds. In fact, unlike motor-vehicle accidents, here both sides are unprotected and at high injury risk. Analogous results can also be found in Savolainen and Mannering (*10*). Another contributing factor to the significance of this collision type could be the high share of motorcycle use in developing countries which accounts for the relatively high frequency of two motor crashes (i.e., 8.1% of the whole motorcycle-involved accident according to **Table 3**).

Collision mode also plays a role in the severity of a crash since head-on accidents are significantly associated with more severe crashes. Compared to the other modes, a crash in this category increases the odds of injury and fatality by 0.001 and 0.0001, respectively. This finding is also supported by evidence from the literature (Shajith et al. (*27*); Savolainen and Mannering (*10*)).

## 5.2 Road Characteristics

The model indicates that the occurrence of a crash in highway segments reduces the probabilities of injury and fatality by 0.039 and 0.003, respectively, whereas increases the probability of PDO by 0.042. Similar findings are also reported by Farid and Ksaibati (*6*). Contrarily, secondary street motorcycle crashes are more likely to be severe, increasing the probability of injury and fatal accidents by 0.01 and 0.001, respectively. Previous research also supports these results (Manan et al. (*7*)). One potential underlying factor could be standard traffic surveillance in highways and limited traffic controls in secondary streets. To shed light on the relationship between traffic controls and risky behaviors, a Chi-square test between road type and human factor variables is conducted. **Figure 3** illustrates the Chi-square residuals between road type categories and rule ignorance (a category of human factor variable) as well as the frequency of each road type (from **Table 3**). the rule ignorance category is associated with crashes in which the motorcycle rider ignores the traffic rules (e.g., helmet, traffic signs, and red-light violation). According to **Figure 3**, crashes occurring in secondary streets have a strong positive correlation with rule ignorance; while highway cases are negatively correlated with rule ignorance. This trend clearly supports the idea that, in our data, motorcycle riders tend to violate the rules in minor and local roadways due to poor traffic control.

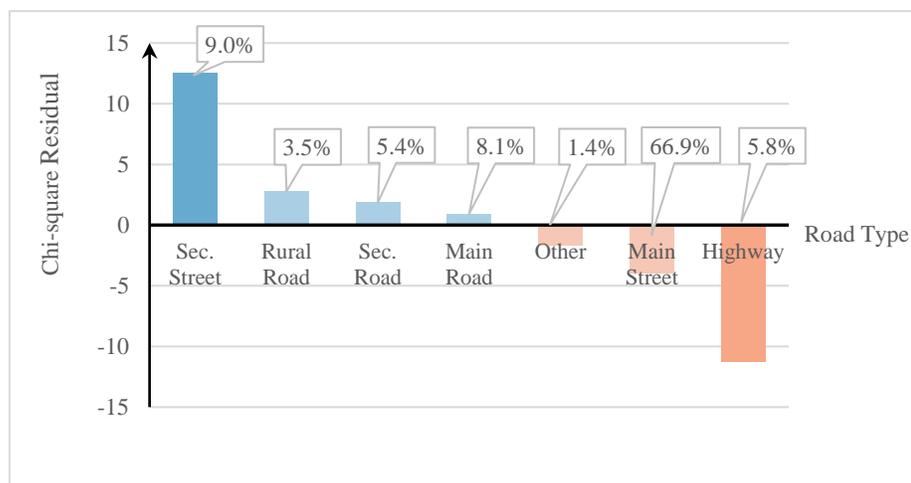

**Figure 3: Chi-square Test Residuals between Road type and Rule Ignorance**
**(Labels above each bar signifies frequency % of each read type)**

**5.3 Temporal Characteristics**

According to the results, spring/summer category affects crash severity by increasing the probability of injury and fatality by 0.044 and 0.003, respectively. This also implies that the base category (i.e., fall/winter) has an inverse effect. This conclusion is in line with Pervez et al. (*25*). A paramount reason for this trend is the different number of motorcycle trips in various seasons. According to **Table 3**, nearly 60% of motorcycle-involved crashes have occurred in warm seasons.

Concerning the time-of-day attribute, night and dawn/dusk time intervals negatively affect injury severity. Happening a crash at night increases the probability of the injury and fatal crashes by 0.043 and 0.003, respectively. Similarly, dawn or dusk variables also contribute to increasing the injury probability by 0.02 and the fatality probability by 0.001. These findings are also confirmed by previous research (*7*, *11*, *25*, *26*), and reportedly, due to the drivers' fatigue and poor lighting level (*39*). Furthermore, the malfunction of motorcycles, specifically the head-light problem, can aggravate this effect which is the direct consequence of non-compulsory maintenance checks and low enforcement in the country.

**5.4 Motorcycle Rider Characteristics**

Three motorcycle rider characteristics in the model have statistically significant impacts on the injury severity: rider's fault status, age, and education level. The base categories for these variables are considered as at-fault, middle-aged, and diploma degree, respectively.

The results suggest that the likelihood of more severe accidents increases when the motorcycle rider is not at fault. More specifically, shifting rider's status from at-fault to not-at-fault would raise the probability of injury and fatal crashes by 0.122 and 0.008, respectively. This result is reasonable considering if a rider is at fault, he/she is more aware of the situation and can react quickly. This also means that in not-at-fault cases, motorcyclists have a higher chance of losing their control and balance due to unexpected conditions and become exposed to more severe damages.

The model also indicates that being a younger driver (i.e., aged less than 25 years) increases the probability of injury and fatal by 0.055 and 0.004, respectively. According to the literature, this finding could be attributed to the prevalence of unskilled riders and the tendency to disobey the rules at this age group (*11*, *40*). Another potential justification is that many younger motorcycle riders may do not have a valid motorcycle license. To test this hypothesis, a Chi-square test between rider age and license status was employed. The test statistics is significant (p<0.001) suggesting that the null hypothesis of no association between these two variables can be rejected. **Figure 4** presents the chi-square test residuals as well as the frequency percentages across different categories of motorcycle age and license type. As shown, the two categories of without license and under 25 are strongly correlated with a positive sign. Moreover, 14.3% of all riders (i.e., 5561/38833) fall into this portion of our data (i.e., under 25 and without a motorcycle license) among which 30% (i.e., 1621/5561) are even under eligible license age (i.e., 18 years old). Unlike young riders, middle-aged riders tend to have a motorcycle license. This correlated pattern between age group and license state is reported by Kashani et al. (*41*) for motorcyclists in Iran. This correlation is also consistent with our previous conclusion regarding the lack of traffic rules enforcement upon motorcycle riders.

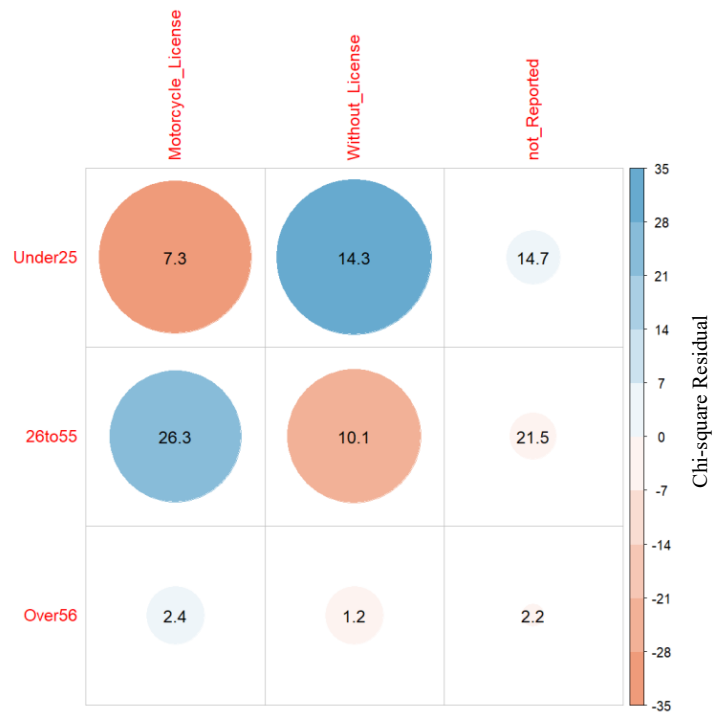

**Figure 4: Chi-square Test Residuals between Rider's Age and Rider's License Status** (Size of bubbles is proportional to residuals and numbers in cells correspond to frequency % among observations)

Being an elderly rider (i.e., over 56) also has a negative impact on injury severity by increasing the probability of injury and fatality by 0.084 and 0.006, respectively. This increase is even more considerable than that of young riders. The decline in vision, cognitive functioning, inability to react quickly are some challenges putting elderly riders at high risk of severe crashes (*42*), especially for the motorcycle riders, that on-time reaction and decision-making is of crucial importance. These results are in good agreement with Pervez et al. (*25*); Quddus et al. (*11*); and Vajari et al. (*26*).

With respect to the education status, by considering the diploma and high school education as the base case, it is observed that riders with education levels under diploma are substantially associated with severe crashes. Belonging to this category engenders an increase in both injury and fatal crash probabilities by 0.017 and 0.001, respectively. Conversely, the negative sign in the college education coefficient implies that crashes in this category tend to be less severe. When a rider has a college or higher education, the probability of PDO increases by 0.076, whereas the probability of injury and fatal crashes reduces by 0.071 and 0.005, respectively. These outcomes depict the positive impact of education on making individuals more disciplined riders. In other words, less educated riders may less adhere to traffic rules (e.g., helmet use, red light rules), as has been emphasized by Waseem et al. (*16*) and Kashani et al. (*41*).

## 6 POLICY IMPLEMENTATION

In this section, upon the results of the presented model, we recommend countermeasures to improve the safety of motorcyclists and reduce the number of motorcycle-involved accidents in developing countries. In this sense, we propose three types of applicable policies, including traffic rule enforcement, educational programs, and road-specific adjustments.

According to the model outcome, there is a wide gap in traffic rule enforcement upon motorcycle riders. The World Health Organization has assigned an enforcement score of 5 out of 10 to Iran in 2018 (*1*). For example, helmet use among motorcyclists of Iran is estimated to be 59% (*1*), whereas this rate in the United States, as a developed country, is at least 71% in the same year (*43*). Further, as underlined in the case of motor-pedestrian crashes, pedestrians are not penalized when violating the traffic rules. It is estimated that nearly 80% of pedestrians in the country do not comply with traffic laws (*44*). These findings highlight the necessity of developing a comprehensive regulation enforcement system for motorcycle riders and pedestrians. According to **Table 5**, the maximum marginal effect of all variables is related to motorcycle-pedestrian crashes (for both injury and fatal crashes). Accordingly, countermeasures targeting both riders and pedestrians would remarkably boost the safety of this group.

As another countermeasure, the government can provide educational programs on assorted platforms, especially targeting young riders who are more likely to have risky riding behaviors. For instance, programs embedded in the licensing procedures can be effective. Licensing system in Iran lacks oriented-based educational programs, especially for motorcyclists, and individual's skills are developed as a result of the self-training process. Therefore, embedding educational programs as a prerequisite part of getting a motorcycle license can be substantially effective in the long run. Similar recommendations are also suggested by Chang and Yeh (*40*) for Taiwan.

Last but not least, we recommend road-specific adjustments to upgrade the safety of motorcyclists and pedestrians. As mentioned in section 5.2, secondary urban road segments are associated with severe crashes. Rider's speeding and unsafe maneuvers due to the lack of traffic control systems may account for this trend. Installing traffic signs can be a general countermeasure to reduce the risk in such roads; however, depending on the condition of a secondary street and the volume of motorcycle users, more serious measures can be considered. For instance, in secondary road segments with high motorcycle volume, a separate exclusive path can be designated to improve the safety of motorcycle riders. Implementation of this strategy in Malaysia has reduced motorcycle deaths by 600% (*45*).

## 7   CONCLUSION

In this paper, we study the characteristics of the severity of motorcycle-involved crashes in Iran, as a developing country. In this regard, we employed Iran's crash records from March 2018 to March 2019. Notwithstanding 20% share of motorcycle-involved crashes from all, 87% of them are injury crashes. However, this ratio for the other crashes is nearly 36%. Moreover, the data statistics suggest that motorcycle-involved crashes make up 5% of PDO, 38% of the injury, and 15% of the fatal crashes. Furthermore, Chi-square test results revealed that the highest association between injury crashes and all vehicle types is related to motorcycles.

To determine the contributing factors to the severity of motorcycle-involved crashes, we utilized the ordered logit model. The results suggest that three collision types (motorcycle-pedestrian, overturn, and two motorcycles) increase the probability of severe crashes. Besides, head-on collision negatively affects severity in the collision angle category. About the road type, collisions taking place in highways are correlated with PDO; whereas secondary street is associated with injury and fatal; crashes. In this regard, using the Chi-square test, we demonstrate that there is a positive correlation between motorcycle riders' rule ignorance (e.g., helmet use, speeding) and urban secondary streets. However, the correlation between highway and rule ignorance is negative. Three temporal characteristics are found to be significant in increasing the likelihood of severe crashes (spring/summer seasons, night, and dawn/dusk hours). Moreover, not-at-fault motorcycle riders are more susceptible to injury and fatal crashes. Moreover, regarding rides' age, under 25 and over 56 aged motorcycle riders are more likely to be involved in severe crashes. As an underlying factor, we discovered a positive association between the under 25 age group and riders without a license, employing Chi-square test residuals. Finally, per the results, riders with secondary school or less education are more likely to be categorized in more severe crashes. In contrast, riders with college degrees decrease the probability of severe crashes.

Based on the model's marginal effects, the most effective attribute on crash severity is collision type comprising two variables motorcycle-pedestrian and overturn. Being involved in a motorcycle-pedestrian crash results in an increase in the probability of injury and fatality by 0.289 and 0.019, respectively. Moreover, overturn motorcycle crashes are more likely to be an injury and fatal crashes by 0.227 and 0.015, respectively. Apart from the aforementioned variables, the maximum increase in injury and fatality probability for all variables is 0.122 and 0.008, respectively.

As per the research findings, we recommended three countermeasures for safety improvement, including traffic rule enforcement, educational programs, and road-specific adjustments. We also highlight the lack of traffic rule enforcement for motorcyclists and pedestrians. The score of traffic law enforcement upon motorcyclists in Iran in 2018 is 5 out of 10 (*1*), which is confirmed by our results.

As a rewarding future extension of our research, we recommend employing more crash severity classes in the context of developing countries. Since data limited our model to consider three injury classes, we were unable to distinguish slight injuries from incapacitating injuries that may have different patterns. This issue is addressed by (*19*, *26*, *31*) for developed countries, usually because of the availability of reach and detailed crash databases.

**Author Contribution Statement**

All the authors contribute equally in all parts of this article.

**REFERENCES**

1.  World Health Organization. Global Status Report on Road Safety 2018. *Geneva, Switzerland: World Health Organization.*

2.  NHTSA. Motorcycle Safety. *National Traffic Highway Safety Administration*, 2021, Retrieved from: https://www.nhtsa.gov/road-safety/.

3.  Lin, M. R., and J. F. Kraus. A Review of Risk Factors and Patterns of Motorcycle Injuries. *Accident Analysis and Prevention*, Vol. 41, No. 4, 2009, pp. 710–722. https://doi.org/10.1016/j.aap.2009.03.010.

4.  IRNA. Occupation of Motorcyclists. *Islamic Republic News Agency*, 2021, Retrieved from: https://www.irna.ir/news.

5.  Safaei, B., N. Safaei, A. Masoud, and S. Seyedekrami. Weighing Criteria and Prioritizing Strategies to Reduce Motorcycle-Related Injuries Using Combination of Fuzzy TOPSIS and AHP Methods. *Advances in transportation studies*, Vol. 54, 2021.

6.  Farid, A., and K. Ksaibati. Modeling Severities of Motorcycle Crashes Using Random Parameters. *Journal of Traffic and Transportation Engineering*, Vol. 8, No. 2, 2021, pp. 225–236. https://doi.org/10.1016/j.jtte.2020.01.001.

7.  Marizwan, M., A. Manan, A. Várhelyi, A. Kemal, and H. Hanis. Road Characteristics and Environment Factors Associated with Motorcycle Fatal Crashes in Malaysia. *IATSS Research*, Vol. 42, No. 4, 2018, pp. 207–220. https://doi.org/10.1016/j.iatssr.2017.11.001.

8.  Shaheed, M. S., and K. Gkritza. A Latent Class Analysis of Single-Vehicle Motorcycle Crash Severity Outcomes. *Analytic Methods in Accident Research*, Vol. 2, 2014, pp. 30–38. https://doi.org/10.1016/j.amar.2014.03.002.

9.  Geedipally, S. R., P. A. Turner, and S. Patil. Analysis of Motorcycle Crashes in Texas with Multinomial Logit Model. *Transportation Research Record*, No. 2265, 2011, pp. 62–69. https://doi.org/10.3141/2265-07.

10. Savolainen, P., and F. Mannering. Probabilistic Models of Motorcyclists' Injury Severities in Single- and Multi-Vehicle Crashes. *Accident Analysis and Prevention*, Vol. 39, No. 5, 2007, pp. 955–963. https://doi.org/10.1016/j.aap.2006.12.016.

11. Quddus, M. A., R. B. Noland, and H. C. Chin. An Analysis of Motorcycle Injury and Vehicle Damage Severity Using Ordered Probit Models. *Journal of safety research*, Vol. 33, No. 4, 2002, pp. 445–462. https://doi.org/10.1016/S0022-4375(02)00051-8.

12. Solagberu, B. A., C. K. P. Ofoegbu, A. A. Nasir, O. K. Ogundipe, A. O. Adekanye, and L. O. Abdur-Rahman. Motorcycle Injuries in a Developing Country and the Vulnerability of Riders, Passengers, and Pedestrians. *Injury Prevention*, Vol. 12, No. 4, 2006, pp. 266–268. https://doi.org/10.1136/ip.2005.011221.

13. Chen, Z., and W. D. Fan. A Multinomial Logit Model of Pedestrian-Vehicle Crash Severity in North Carolina. *International Journal of Transportation Science and Technology*, 2018. https://doi.org/10.1016/j.ijtst.2018.10.001.

14. Sohrabi, S., A. Khodadadi, S. M. Mousavi, B. Dadashova, and D. Lord. Quantifying the Automated Vehicle Safety Performance: A Scoping Review of the Literature, Evaluation of Methods, and Directions for Future Research. *Accident Analysis & Prevention*, Vol. 152, 2021, p. 106003. https://doi.org/https://doi.org/10.1016/j.aap.2021.106003.

15. Lin, M. R., S. H. Chang, W. Huang, H. F. Hwang, and L. Pai. Factors Associated with Severity of Motorcycle Injuries among Young Adult Riders. *Annals of Emergency Medicine*, Vol. 41, No. 6, 2003, pp. 783–791. https://doi.org/10.1067/mem.2003.186.

16. Waseem, M., A. Ahmed, and T. Usman. Factors a Ff Ecting Motorcyclists ' Injury Severities : An Empirical Assessment Using Random Parameters Logit Model with Heterogeneity in Means and Variances. *Accident Analysis and Prevention*, Vol. 123, 2019, pp. 12–19. https://doi.org/10.1016/j.aap.2018.10.022.

17. Das, S., A. Dutta, K. Dixon, L. Minjares-Kyle, and G. Gillette. Using Deep Learning in Severity Analysis of At-Fault Motorcycle Rider Crashes. *Transportation Research Record*, Vol. 2672, No. 34, 2018, pp. 122–134. https://doi.org/10.1177/0361198118797212.

18. Das, S., I. Tsapakis, and A. Khodadadi. Safety Performance Functions for Low-Volume Rural Minor Collector Two-Lane Roadways. *IATSS Research*, No. xxxx, 2021. https://doi.org/10.1016/j.iatssr.2021.02.004.

19. Jones, S., S. Gurupackiam, and J. Walsh. Factors Influencing the Severity of Crashes Caused by Motorcyclists: Analysis of Data from Alabama. *Journal of Transportation Engineering*, Vol. 139, No. 9, 2013, pp. 949–956. https://doi.org/10.1061/(ASCE)TE.1943-5436.0000570.

20. Wali, B., A. J. Khattak, and N. Ahmad. Examining Correlations between Motorcyclist's Conspicuity, Apparel Related Factors and Injury Severity Score: Evidence from New Motorcycle Crash Causation Study. *Accident Analysis and Prevention*, Vol. 131, No. April, 2019, pp. 45–62. https://doi.org/10.1016/j.aap.2019.04.009.

21. Haque, M., H. C. Chin, and A. K. Debnath. AN INVESTIGATION ON MULTI-VEHICLE MOTORCYCLE CRASHES USING LOG-LINEAR MODELS. *Safety Science*, Vol. 50, No. 2, 2012, pp. 352–362.

22.    Halbersberg, D., and B. Lerner. Young Driver Fatal Motorcycle Accident Analysis by Jointly Maximizing Accuracy and Information. *Accident Analysis and Prevention*, Vol. 129, No. August 2018, 2019, pp. 350–361. https://doi.org/10.1016/j.aap.2019.04.016.

23.    Shankar, V., and F. Mannering. An Exploratory Multinomial Logit Analysis of Single-Vehicle Motorcycle Accident Severity. *Journal of Safety Research*, Vol. 27, No. 3, 1996, pp. 183–194. https://doi.org/10.1016/0022-4375(96)00010-2.

24.    Wahab, L., and H. Jiang. A Comparative Study on Machine Learning Based Algorithms for Prediction of Motorcycle Crash Severity. *PLoS ONE*, Vol. 14, No. 4, 2019, pp. 1–17. https://doi.org/10.1371/journal.pone.0214966.

25.    Pervez, A., J. Lee, and H. Huang. Identifying Factors Contributing to the Motorcycle Crash Severity in Pakistan. *Journal of Advanced Transportation*, Vol. 2021, 2021. https://doi.org/10.1155/2021/6636130.

26.    Abrari Vajari, M., K. Aghabayk, M. Sadeghian, and N. Shiwakoti. A Multinomial Logit Model of Motorcycle Crash Severity at Australian Intersections. *Journal of Safety Research*, Vol. 73, 2020, pp. 17–24. https://doi.org/10.1016/j.jsr.2020.02.008.

27.    Shajith, S. L. A., H. R. Pasindu, and R. K. T. K. Ranawaka. Evaluating the Risk Factors in Fatal Accidents Involving Motorcycle – Case Study on Motorcycle Accidents in Sri Lanka. *Engineer: Journal of the Institution of Engineers, Sri Lanka*, Vol. 52, No. 3, 2019, p. 33. https://doi.org/10.4038/engineer.v52i3.7363.

28.    De Oliveira, A. L., A. Petroianu, D. M. V. Gonçalves, G. A. Pereira, and L. R. Alberti. Characteristics of Motorcyclists Involved in Accidents between Motorcycles and Automobiles. *Revista da Associacao Medica Brasileira*, Vol. 61, No. 1, 2015, pp. 61–64. https://doi.org/10.1590/1806-9282.61.01.061.

29.    Chichom-Mefire, A., J. Atashili, J. G. Tsiagadigui, C. Fon-Awah, and M. Ngowe-Ngowe. A Prospective Pilot Cohort Analysis of Crash Characteristics and Pattern of Injuries in Riders and Pillion Passengers Involved in Motorcycle Crashes in an Urban Area in Cameroon: Lessons for Prevention. *BMC Public Health*, Vol. 15, No. 1, 2015, pp. 1–8. https://doi.org/10.1186/s12889-015-2290-4.

30.    Rifaat, S. M., R. Tay, and A. De Barros. Severity of Motorcycle Crashes in Calgary. *Accident Analysis and Prevention*, Vol. 49, No. July 2015, 2012, pp. 44–49. https://doi.org/10.1016/j.aap.2011.02.025.

31.    Eustace, D., V. K. Indupuru, and P. Hovey. Identification of Risk Factors Associated with Motorcycle-Related Fatalities in Ohio. *Journal of Transportation Engineering*, Vol. 137, No. 7, 2011, pp. 474–480. https://doi.org/10.1061/(ASCE)TE.1943-5436.0000229.

32. Oginni, F. O., S. O. Ajike, O. N. Obuekwe, and O. Fasola. A Prospective Multicenter Study of Injury Profile, Severity and Risk Factors in 221 Motorcycle-Injured Nigerian Maxillofacial Patients. *Traffic Injury Prevention*, Vol. 10, No. 1, 2009, pp. 70–75. https://doi.org/10.1080/15389580802496968.

33. Chang, H. L., and T. H. Yeh. Risk Factors to Driver Fatalities in Single-Vehicle Crashes: Comparisons between Non-Motorcycle Drivers and Motorcyclists. *Journal of Transportation Engineering*, Vol. 132, No. 3, 2006, pp. 227–236. https://doi.org/10.1061/(ASCE)0733-947X(2006)132:3(227).

34. P. Iamtrakul, K. Hokao, Tanaboriboon, Y. Analysis of Motorcycle Accidents in Developing Countries: A Case Study of Khon Kaen, Thailand. *Journal of the Eastern Asia Society for Transportation Studies*, Vol. 5, No. December 2014, 2003, pp. 147–162.

35. Donald, S. Your Chi-Square Test Is Statistically Significant: Now What? *Practical Assessment, Research and Evaluation*, Vol. 20, No. 8, 2015, pp. 1–10.

36. Eluru, N., and S. Yasmin. A Note on Generalized Ordered Outcome Models. *Analytic Methods in Accident Research*, Vol. 8, 2015, pp. 1–6. https://doi.org/10.1016/j.amar.2015.04.002.

37. Chen, C., G. Zhang, H. Huang, J. Wang, and R. A. Tarefder. Examining Driver Injury Severity Outcomes in Rural Non-Interstate Roadway Crashes Using a Hierarchical Ordered Logit Model. *Accident Analysis and Prevention*, Vol. 96, 2016, pp. 79–87. https://doi.org/10.1016/j.aap.2016.06.015.

38. Train, K. E. *Discrete Choice Methods with Simulation*. 2003.

39. Rowden, P., D. Steinhardt, and M. Sheehan. Road Crashes Involving Animals in Australia. *Accident Analysis and Prevention*, Vol. 40, No. 6, 2008, pp. 1865–1871. https://doi.org/10.1016/j.aap.2008.08.002.

40. Chang, H. L., and T. H. Yeh. Motorcyclist Accident Involvement by Age, Gender, and Risky Behaviors in Taipei, Taiwan. *Transportation Research Part F: Traffic Psychology and Behaviour*, Vol. 10, No. 2, 2007, pp. 109–122. https://doi.org/10.1016/j.trf.2006.08.001.

41. Kashani, A. T., M. M. Besharati, and A. Mohamadian. Analyzing Motorcycle Crash Pattern and Riders ' Fault Status at a National Level : A Case Study from Iran. *International Journal of Transportation Engineering*, Vol. 5, No. 1, 2017, pp. 87–101.

42. CDC. Older Adult Drivers. *Centers for Disease Control and Prevention*, 2020, p. Retrieved from: https://www.cdc.gov/transportation.

43. National Highway Traffic Safety Administration. Motorcycle Helmet Use in 2019 - Overall Results. *U.S. Department of Transportation*, 2020, p. 5.

44.  Aghdam, F. B., H. S. Bazargani, P. Sarbakhsh, T. Pashaie, K. Ponent, and M. Nicknejad. Pedestrians in Iran : Determinants of Unsafe Traffic Behaviors of Pedestrians. *Research Square*, pp. 1–11.

45.  Radin Umar, R. S. Motorcycle Safety Programmes in Malaysia: How Effective Are They? *International journal of injury control and safety promotion*, Vol. 13, No. 2, 2006, pp. 71–79. https://doi.org/10.1080/17457300500249632.